\documentclass[twocolumn,showpacs,preprintnumbers,superscriptaddress,amsmath,amssymb]{revtex4}

\usepackage{graphicx}
\usepackage{dcolumn}
\usepackage{bm}

\usepackage{color}

\begin{document}


\preprint{1}

\title{Nuclear cluster  structure effect on elliptic and triangular flows in  heavy-ion collisions }

\author{S. Zhang}
\affiliation{Shanghai Institute of Applied Physics, Chinese Academy of Sciences, Shanghai 201800, China}
\author{Y. G. Ma\footnote{Author to whom all correspondence should be addressed: ygma@sinap.ac.cn}}
\affiliation{Shanghai Institute of Applied Physics, Chinese Academy of Sciences, Shanghai 201800, China}
\affiliation{University of Chinese Academy of Sciences, Beijing 100049, China}
\affiliation{ShanghaiTech University, Shanghai 200031, China}
\author{J. H. Chen}
\affiliation{Shanghai Institute of Applied Physics, Chinese Academy of Sciences, Shanghai 201800, China}
\author{W. B. He}
\affiliation{Institute of Modern Physics, Fudan University, Shanghai 200433, China}
\author{C. Zhong}
\affiliation{Shanghai Institute of Applied Physics, Chinese Academy of Sciences, Shanghai 201800, China}

\date{\today}

\begin{abstract}
The initial geometry effect on collective flows, which are inherited from initial projectile structure, is studied in relativistic heavy-ion collisions of $^{12}$C + Au by using a multi-phase transport model (AMPT). Elliptic flow ($v_2$) and triangular flow ($v_3$) which are significantly resulted from the chain and triangle structure of $^{12}\mathrm{C}$ with three-$\alpha$ clusters, respectively, in central $^{12}\mathrm{C}$+$^{197}\mathrm{Au}$ collisions are compared with the flow from  the Woods-Saxon distribution of nucleons in $^{12}\mathrm{C}$. $v_3/v_2$ is proposed as a probe to distinguish the pattern of $\alpha$-clustered $^{12}\mathrm{C}$. This study demonstrates that the initial geometry of the collision zone inherited from nuclear structure can be explored by collective flow at the final stage in heavy-ion collisions.
\end{abstract}

\maketitle

\section{Introduction}

Heavy-ion collision at relativistic energy provides abundant information about nuclear or partonic matter by the final state products~\cite{RHICWhitePaper-BRAHMS, RHICWhitePaper-PHOBOS, RHICWhitePaper-STAR, RHICWhitePaper-PHENIX, FisrtResultsALICE}. The observables, such as collective flow~\cite{STARv1BES,STARv2BES, STARv3,STARFlowCME,Song}, Hanbury-Brown-Twiss (HBT) correlation~\cite{STAR-HBTBES,PHENIX-HBTBES,Lacey,ZhangWN}, chiral  electric-magnetic effects \cite{Dima,HuangXG,MaGL,DengWT} and fluctuation~\cite{STAR-FLUCBES,Ko,Luo}, are sensitive to not only properties of the hot-dense matter but also initial state of pre-collision system. The initial geometrical fluctuation can significantly contribute on collective flow, which attracts  some theoretical attentions. Hydrodynamical model~\cite{InitGeoHydo-1,InitGeoHydo-2, InitGeoHydo-3, InitGeoHydo-4,Song,hydro2} can describe the initial state fluctuation event by event and address its effect on collective flow and viscosity. AMPT model~\cite{AMPTInitFluc-1,AMPTInitFluc-2,AMPTInitFluc-3} can present initial geometry fluctuations of partons created in Au+Au collisions and give the quantitative description of the elliptic and triangular flows. And other works~\cite{PartPlane-1,PartPlane-2,PartPlane-3} contribute a lot to flow/eccentricity analysis method related to initial geometry fluctuations.

It was proposed that the intrinsic geometry of light nuclei can be studied in heavy-ion collisions by exploring elliptic and triangular flow formed in the collisions~\cite{AlphaClusterHIC-1,AlphaClusterHIC-2}. In these works the carbon is considered with 3-$\alpha$ and collides against a heavy nucleus at very high energies. Their calculation showed significant quantitative and qualitative differences between the $\alpha$-clustered and uniform $^{12}\mathrm{C}$ nucleus occur in some quantities such as the triangular flow and its event-by-event fluctuations, or the correlations of the elliptic and triangular flows. Since the $\alpha$ cluster model was proposed by Gamow~\cite{GamowAlphaModel}, $\alpha$-clustering light nuclei  have been studied for more than four decades~\cite{AlphaCluster-1,AlphaCluster-2,Kanada} and experimental evidences for clustering from fragmentation can be found in Ref.~\cite{AlphaCluster-3}. Light $4N$ nuclei from $^{12}\mathrm{C}$ to $^{28}\mathrm{Si}$ were investigated in the $\alpha$-clustering model of Margenau, Bloch and Brink~\cite{AlphaModelBRINK}. Recently it was proposed that the giant dipole resonance ~\cite{AlphaModelHe1,AlphaModelHe2} and photonuclear reaction \cite{Huang1,Huang2}  can be taken as useful probes of $\alpha$-clustering configurations in $^{12}\mathrm{C}$ and $^{16}\mathrm{O}$ by some of the authors. Many theoretical studies suggest that  $^{12}\mathrm{C}$ could exhibit triangular or chain distribution of three $\alpha$ clusters and $^{16}\mathrm{O}$ can emerge kite, chain, square or tetrahedron arrangements of four $\alpha$ clusters.

In this work, elliptic and triangular flow are studied in $^{12}\mathrm{C}$ + $^{197}\mathrm{Au}$ collisions at $\sqrt{s_{NN}}$ = 10 and 200 GeV by using the AMPT model~\cite{AMPT2005}. The initial nucleon distribution in $^{12}\mathrm{C}$ and $^{197}\mathrm{Au}$ are initialized by Woods-Saxon shapes from the HIJING model~\cite{HIJING-1,HIJING-2} in original version of AMPT. And $^{12}\mathrm{C}$ is configured with three-$\alpha$ clusters arranged in chain or triangle structure by using the results from an extended quantum molecular dynamics model (EQMD)~\cite{AlphaModelHe1,AlphaModelHe2,Huang1} for investigating the effect on collective flow from initial geometrical distribution of nucleons, i.e. probing the effects of $\alpha$-clusters.

The paper is organised in following. Section II gives the model and methodology for briefly introducing the AMPT and EQMD models as well as the flow calculation methods. Results and discussion is given in Section III where the elliptic ($v_2$) and triangle ($v_3$) flows are discussed in different collision stages, and the coordinate eccentricities  ($\epsilon_n$) are demonstrated. Furthermore, the transformation coefficient from the coordinate anisotropy to momentum anisotropy is presented, and the ratio of $v_3/v_2$ is proposed as a probe for clustering structure. Finally, the conclusion is drawn in Section IV. 

\section{model and methodology}

A multi-phase transport model (AMPT)~\cite{AMPT2005} is employed to study the effect on collective flow from the initial  geometry distribution of nucleons. AMPT has been extensively applied to simulate heavy-ion collisions in a wide colliding energy range from the SPS to the LHC, which contains the initial conditions from the HIJING model~\cite{HIJING-1,HIJING-2}, partonic interactions modelled by a parton cascade model (ZPC)~\cite{ZPCModel}, hadronization in a simple quark coalescence model, and hadronic rescattering simulated by a relativistic transport (ART) model~\cite{ARTModel}. AMPT is  successful to describe physics in relativistic heavy-ion collision for the RHIC~\cite{AMPT2005} and the LHC energies~\cite{AMPTGLM2016}, including pion-pair correlations~\cite{AMPTHBT}, di-hadron azimuthal correlations~\cite{AMPTDiH} as well as  collective flow~etc  \cite{STARFlowAMPT,AMPTFlowLHC}.

\begin{figure*}[htbp]
\includegraphics[width=17cm]{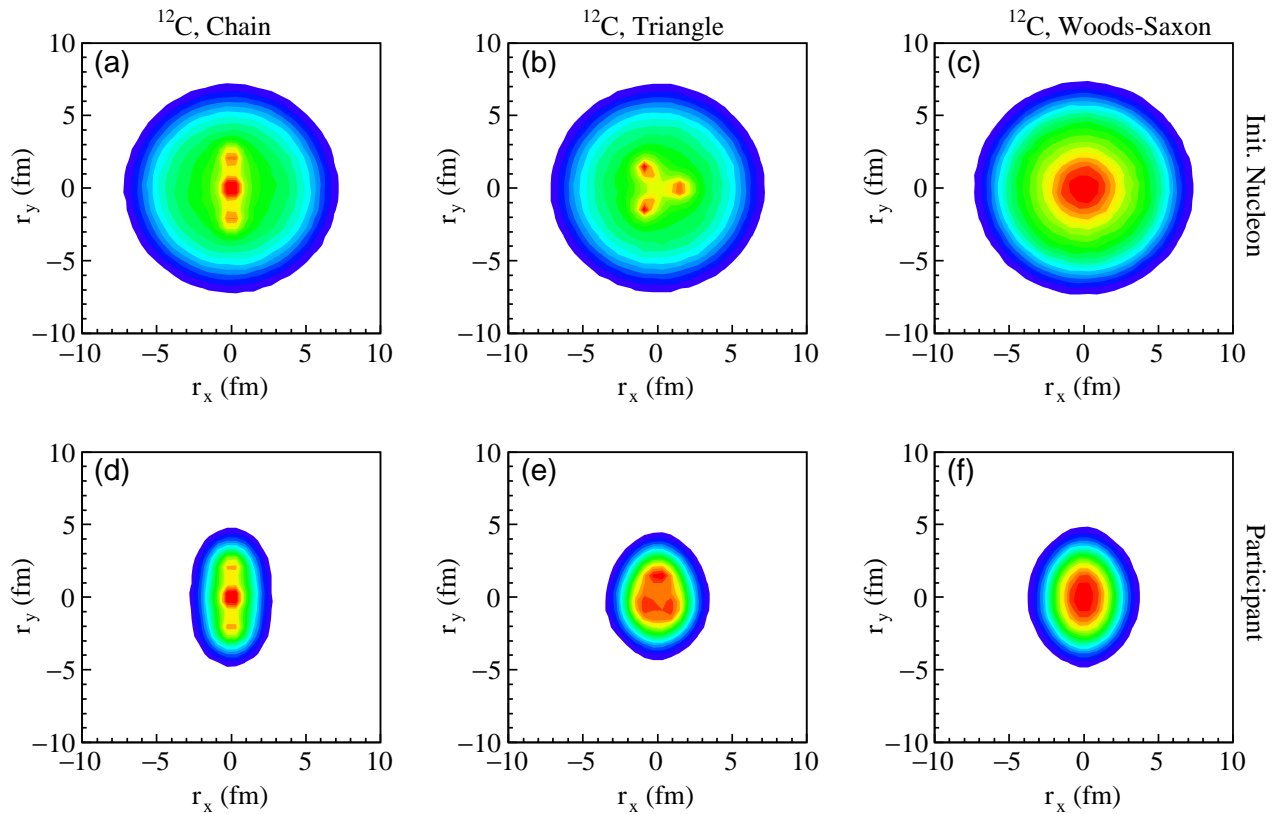}
\caption{
\label{fig:InitNuclPart}(color online)
 In  the most central collisions, initial intrinsic nucleon distribution of  the $^{12}\mathrm{C}$+$^{197}\mathrm{Au}$ system (Upper panels), and the  participant distributions (Lower panels):   (a) and (d) for the $^{12}$C with the Chain $\alpha$-clustering  structure, (b) and (e) for the Triangle  $\alpha$-clustering  structure,   (c) and (f) for  the Woods-Saxon nucleon distribution. 
The nucleon distribution in $^{197}\mathrm{Au}$ always take the Woods-Saxon form.
 }
\end{figure*}

On the other hand, the EQMD model ~\cite{EQMD1996,EQMD1998} was extended from the Quantum Molecular dynamics (QMD) \cite{Aichelin} type models, which have  successfully described physics in low energy heavy-ion reaction, eg. collective  flows, multifragmentation, HBT correlation, nuclear stopping, shear viscosity, giant resonances and pion azimuthal asymmetry \cite{Yan,Ma_Shen,Zhang_Ra,Ma_HBT,Zhou,TaoC,FengZQ,FengZQ2,WangTT}.  Due to an effective Pauli potential is employed, the EQMD model can give reasonable $\alpha$-cluster configurations for $4N$ nuclei. All the $\alpha$-cluster configurations are different by their local minima states, with low excited energy. The parameters for configuring the $\alpha$-clustered $^{12}\mathrm{C}$ are inherited from the EQMD calculation~\cite{AlphaModelHe1,AlphaModelHe2,Huang1}. The initial nucleon distribution in $^{12}\mathrm{C}$ is configured for three cases, as shown in Fig.~\ref{fig:InitNuclPart}: (a) three $\alpha$ clusters in Chain structure, (b) three $\alpha$ clusters in Triangle structure, and (c) nucleons in Woods-Saxon distribution from HIJING model~\cite{HIJING-1,HIJING-2} (Woods-Saxon). The distribution of radial centre of the $\alpha$ clusters in $^{12}\mathrm{C}$ is assumed to be a Gaussian function, $e^{-0.5\left(\frac{r-r_c}{\sigma_{r_c}}\right)^{2}}$, here $r_c$ is average radial center of an $\alpha$ cluster and $\sigma_{r_c}$ is the width of the distribution. And the nucleon inside each  $\alpha$ cluster will be given by Woods-Saxon distribution. The parameters of $r_c$ and $\sigma_{r_c}$ can be obtained from the EQMD calculation~\cite{AlphaModelHe1,AlphaModelHe2,Huang1}.  For  the Triangle structure,   $r_c$ = 1.8 fm and $\sigma_{r_c}$  = 0.1 fm. For Chain structure, $r_c$ = 2.5 fm, $\sigma_{r_c}$  = 0.1 fm for two $\alpha$ clusters and the other one will be at the center in $^{12}\mathrm{C}$. Once the  radial centre of the $\alpha$ cluster is determined, the centers of the three clusters will be placed in an equilateral triangle for the Triangle structure or in a line for Chain structure. For the nucleons in $^{197}\mathrm{Au}$, we just take Woods-Saxon distribution from the HIJING model~\cite{HIJING-1,HIJING-2} (Woods-Saxon). Figure~\ref{fig:InitNuclPart} (a), (b) and (c) show the three cases of intrinsic initial nucleon distribution of the $^{12}\mathrm{C}$+$^{197}\mathrm{Au}$ system in the most central collisions. And the
intrinsic participant distributions are shown in Fig.~\ref{fig:InitNuclPart} (d), (e) and (f), respectively, for the Chain, Triangle and Woods-Saxon distribution initializied $^{12}\mathrm{C}$ in $^{12}\mathrm{C}$+$^{197}\mathrm{Au}$ collisions in the AMPT model. From Fig.~\ref{fig:InitNuclPart} (d), (e) and (f), we can see that the participant distribution inherits the geometrical shape of the initializied $^{12}\mathrm{C}$. The coordinate asymmetry will be transferred to momentum space asymmetry in the evolution of the fireball created in the collisions from hydrodynamical theory~\cite{FlowHydro-1,FlowHydro-2}.

\begin{figure}[htbp]
\includegraphics[width=8.6cm]{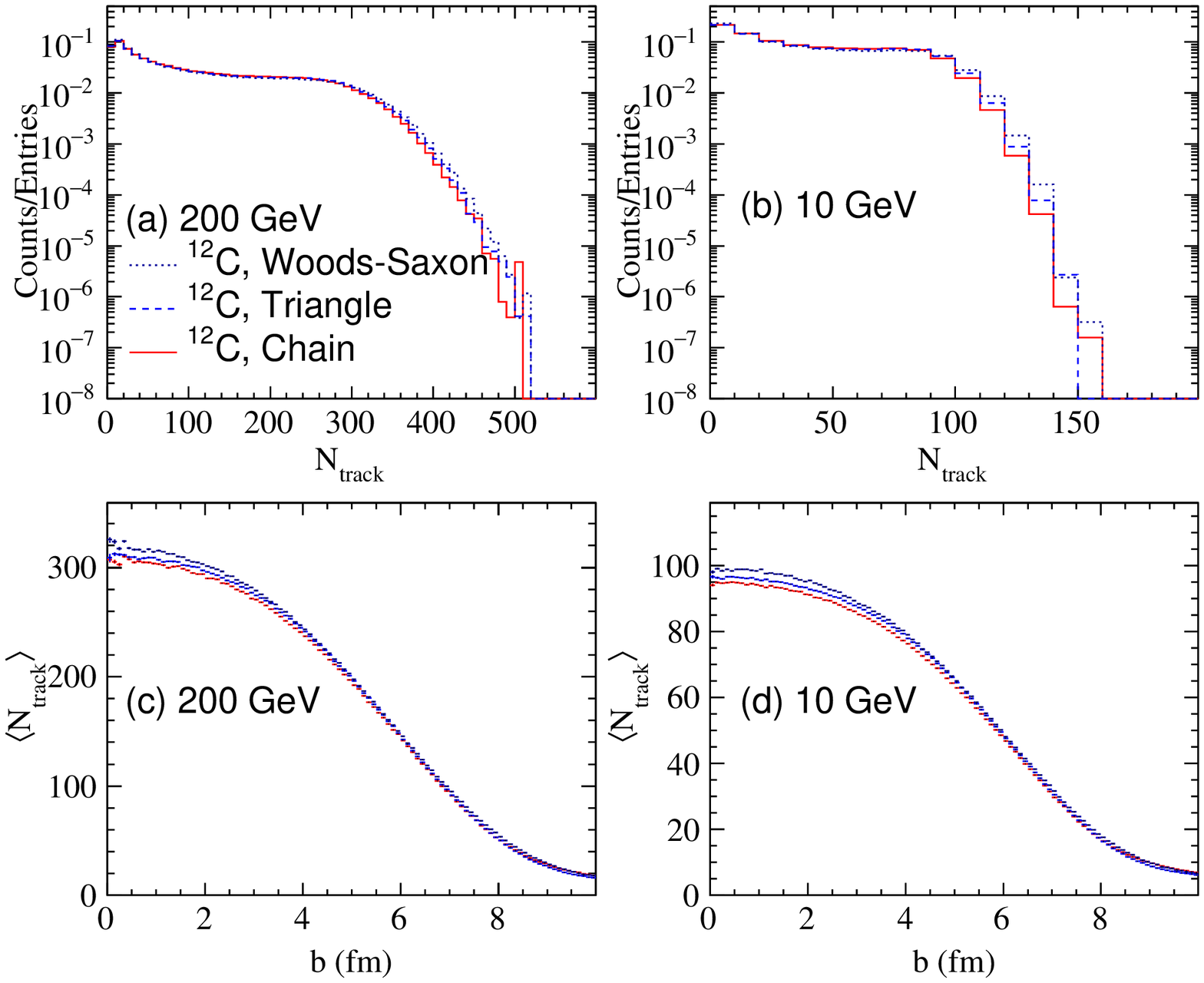}
\caption{
\label{fig:NTrackDst_b}(color online)
The $\mathrm{N_{track}}$ distribution in $^{12}\mathrm{C}$+$^{197}\mathrm{Au}$ collisions at $\sqrt{s_{NN}}$ = 200 GeV (a)  and  10 GeV (b), respectively; Average $\mathrm{N_{track}}$ ($\left<\mathrm{N_{track}}\right>$) as a function of impact parameter $b$ at (c) 200 GeV and (d) 10 GeV, respectively.
 }
\end{figure}

With these intrinsic nucleon distributions of each configuration,  orientation of all nucleons  are rotated randomly for each event in heavy ion collisions.  After accumulating enough events for statistics, the observables are obtained.
Figure~\ref{fig:NTrackDst_b} (a) and (b) present the distribution of number of tracks (particles) ($\mathrm{N_{track}}$) in collisions for the three $^{12}$C configuration cases, namely the Woods-Saxon, Triangle and Chain structures. Noting that the $\mathrm{N_{track}}$ is calculated in the rapidity window ($-2<y<2$) and transverse momentum window ($0.2<p_T<6$) GeV/c for the charged pions ($\pi^{\pm}$), Kaons ($K^{\pm}$) and protons ($p$ and $\bar{p}$). Figure~\ref{fig:NTrackDst_b} (c) and (d) show average $\mathrm{N_{track}}$ ($\left<\mathrm{N_{track}}\right>$) as a function of impact parameter $b$
of $^{12}\mathrm{C}$ and $^{197}\mathrm{Au}$. From Fig.~\ref{fig:NTrackDst_b}  (a) and (c),
and  (b) and (d), the most central collisions can be estimated in impact parameter $b<2$ fm with $\mathrm{N_{track}} > 280$ at $\sqrt{s_{NN}}$ = 200 GeV and $\mathrm{N_{track}} > 90$ at $\sqrt{s_{NN}}$ = 10 GeV.  From the figure, it can be seen that there is no significant different for total track numbers among  three cases for $^{12}\mathrm{C}$ initialization.

The collective properties in heavy-ion collisions can be investigated by the azimuthal anisotropy of detected particles~\cite{Collectivity1992}. The final-state particle azimuthal distribution can be expanded as~\cite{flowMethod1998},
\begin{eqnarray}
E\frac{d^3N}{d^3p} = \frac{1}{2\pi}\frac{d^2N}{p_Tdp_Tdy}\left(1+\sum_{i=1}^{N}2v_n\cos[n(\phi-\Psi_{RP})]\right),
\label{FlowExpansion}
\end{eqnarray}
where $E$ is the energy, $p_T$ is transverse momentum, $y$ is rapidity, and $\phi$ is  azimuthal angle of the particle. $\Psi_{RP}$ is the reaction plane angle. The Fourier coefficients $v_n (n = 1, 2, 3, ...)$ represent collective flow of different orders in azimuthal anisotropies with the form,
\begin{eqnarray}
v_n = \left<\cos(n[\phi-\Psi_{RP}])\right>,
\label{FlowDef}
\end{eqnarray}
where the bracket $\left<\right>$ denotes statistical averaging over particles and events. Harmonic flow $v_n$ can also be calculated through participant plane angle $\Psi_n\{PP\}$ instead of reaction plane $\Psi_{RP}$. In participant coordinate system, the participant plane angle $\Psi_n\{PP\}$ can be obtained event by event using the following equation~\cite{AMPTInitFluc-1,WangJ,PartPlane-1,PartPlane-2,PartPlane-3},
\begin{eqnarray}
\Psi_n\{PP\} = \frac{\mathrm{atan2}\left(\left<r^2\sin\left(n\phi_{P}\right)\right>,\left<r^2\cos\left(n\phi_{P}\right)\right>\right)+\pi}{n},
\label{PartPlanDef}
\end{eqnarray}
where, $\Psi_n\{PP\}$ is the {\it n}th-order participant plane angle, $r$ and $\phi_{P}$ are coordinate position and azimuthal angle of participants in
 the collision zone at initial state, and the average $\left<\cdots\right>$ denotes density weighting.
Then the harmonic flow coefficients with respect to participant plane can  be defined as,
\begin{eqnarray}
v_n\equiv\left<\cos(n[\phi-\Psi_n\{PP\}])\right>.
\label{FlowPPDef}
\end{eqnarray}

Also, the initial coordinate anisotropies can be quantified as participant eccentricity coefficients~\cite{AMPTInitFluc-1,PartPlane-1,PartPlane-2,PartPlane-3},
\begin{eqnarray}
\epsilon_n\{PP\}\equiv\frac{\sqrt{\left<r^2\cos\left(n\phi_{P}\right)\right>^2+\left<r^2\sin\left(n\phi_{P}\right)\right>^2}}{\left<r^2\right>}.
\label{EpsilonPPDef}
\end{eqnarray}

The collisions are initializied by the HIJING model~\cite{HIJING-1,HIJING-2} in AMPT model. Since the initial fluctuation should be taken into account, the initial coordinates to calculate Eq.~(\ref{FlowPPDef}) and~(\ref{EpsilonPPDef}) will be provided by the HIJING process in this work.

This method (participant plane method) has been used to calculate collective flow in different models~\cite{InitGeoHydo-1,InitGeoHydo-2,InitGeoHydo-3,InitGeoHydo-4,AMPTInitFluc-1,AMPTInitFluc-2,AMPTInitFluc-3,PartPlane-1,PartPlane-2,PartPlane-3}.  And it was always applied to discuss the initial geometric fluctuation effect on collective flow because the participant plane angle $\Psi_n\{PP\}$ is constructed by initial energy distribution in coordinate space with the event-by-event fluctuation effects. In this work, the event-by-event participant plane method for flow calculation is employed to calculate elliptic flow $v_2$ and triangular flow $v_3$ in $^{12}\mathrm{C}$ + $^{197}\mathrm{Au}$ collisions at $\sqrt{s_{NN}}$ = 10 GeV and 200 GeV in the AMPT model. And a probe to distinguish different $\alpha$-clustering structure of $^{12}\mathrm{C}$ will be proposed from the results of elliptic flow $v_2$ and triangular flow $v_3$.

\section{Results and discussion}

\begin{figure*}[htbp]
\includegraphics[width=17cm]{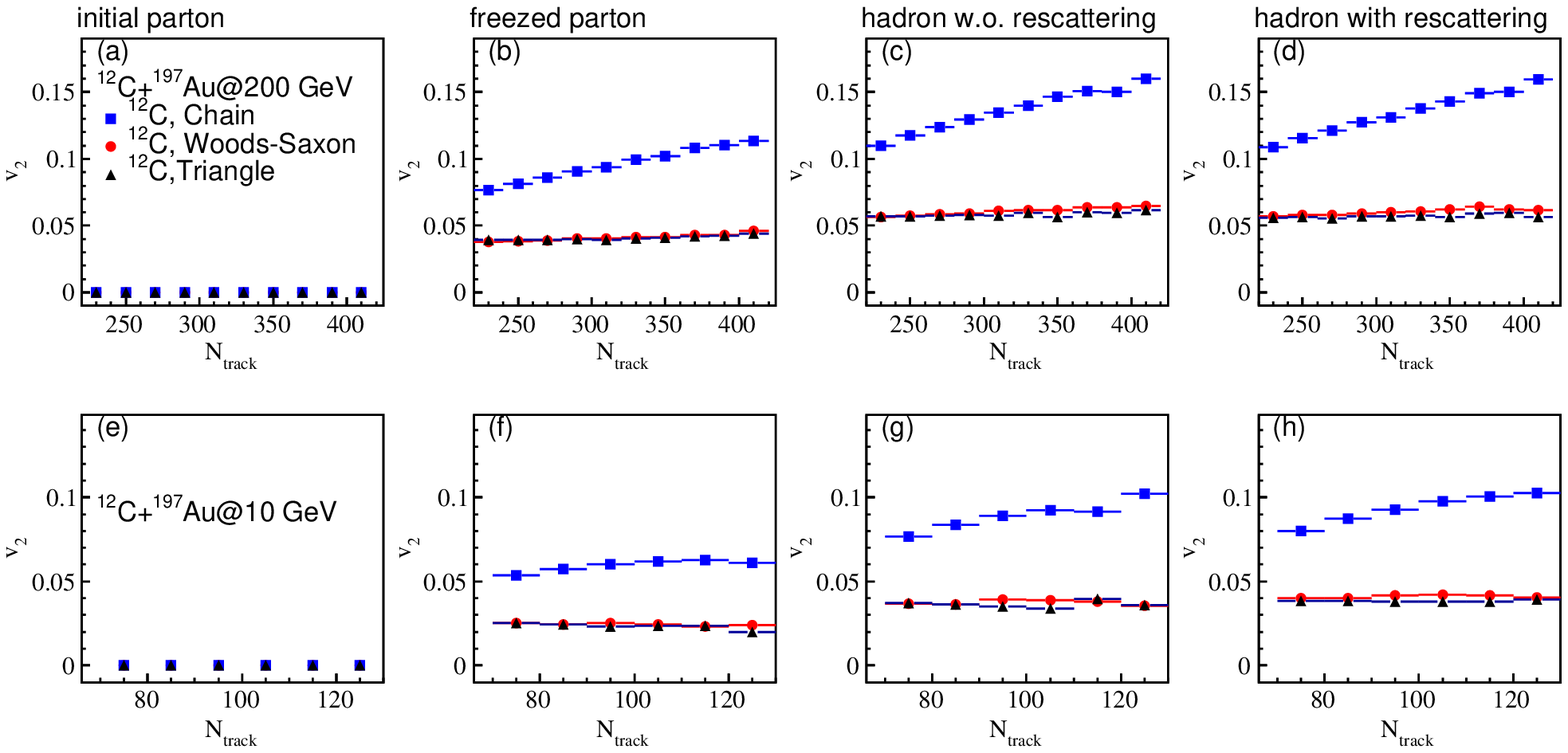}
\caption{
\label{fig:v2Evolution}(color online)
In $^{12}\mathrm{C}$+$^{197}\mathrm{Au}$ collisions at $\sqrt{s_{NN}}$ = 200 GeV (upper panels) and 10 GeV (lower panels), $v_2$ as function of $\mathrm{N_{track}}$: (a) and (e) for  initial partons, (b) and (f) for the freeze-out partons, (c) and (g) for the hadrons without hadronic rescattering, and (d) and (h) for the final-state hadrons.
 }
\end{figure*}

\begin{figure*}[htbp]
\includegraphics[width=17cm]{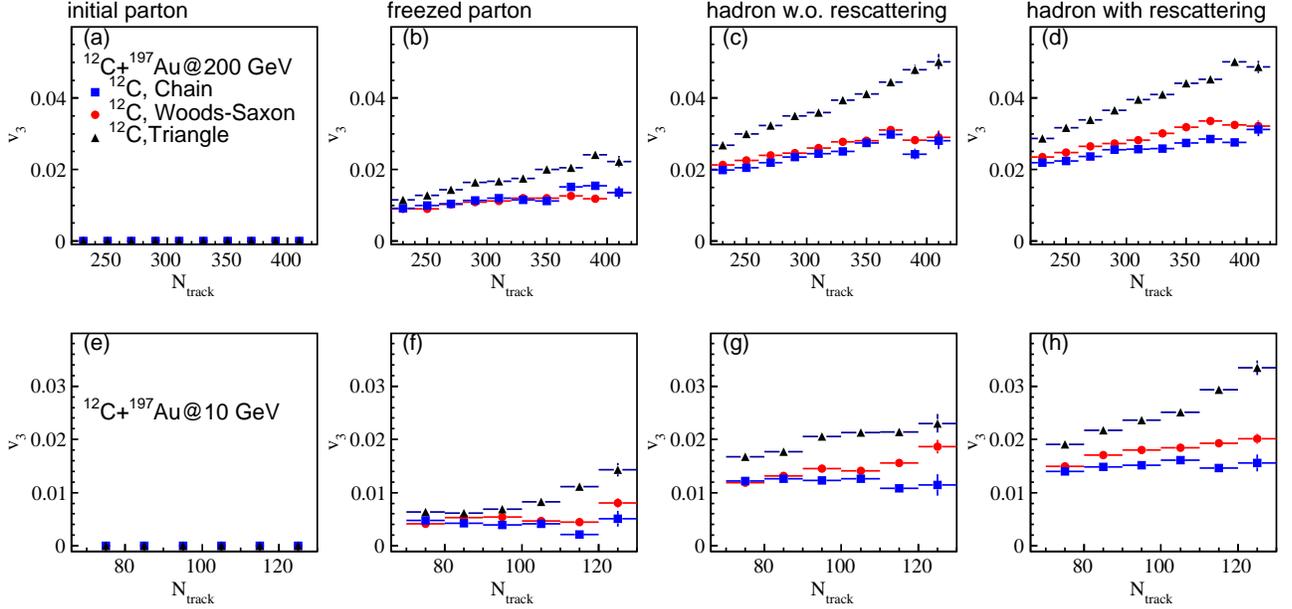}
\caption{
\label{fig:v3Evolution}(color online) Same as Fig.~\ref{fig:v2Evolution} but for  $v_3$.
 }
\end{figure*}

\begin{figure}[htbp]
\includegraphics[width=8.6cm]{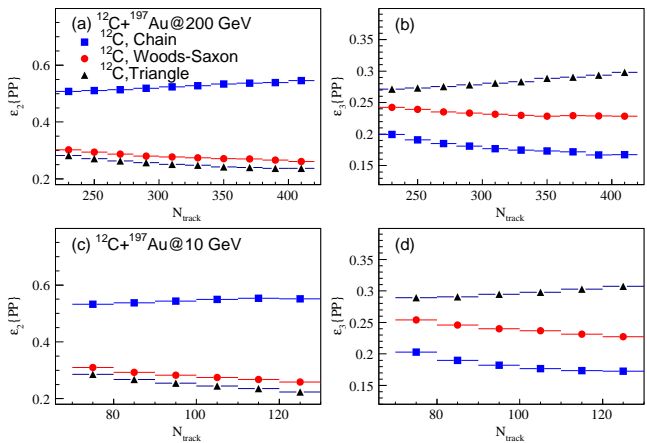}
\caption{
\label{fig:epsilonVn_cent}(color online)
 In $^{12}\mathrm{C}$+$^{197}\mathrm{Au}$ collisions at $\sqrt{s_{NN}}$ = 200 GeV (upper panels) and 10 GeV (lower panels),  second and third order participant eccentricity coefficients, (a) and (c) for $\epsilon_2\{PP\}$ ,  (b) and (d) for $\epsilon_3\{PP\}$, as function of $\mathrm{N_{track}}$.
 }
\end{figure}

\begin{figure}[htbp]
\includegraphics[width=8.6cm]{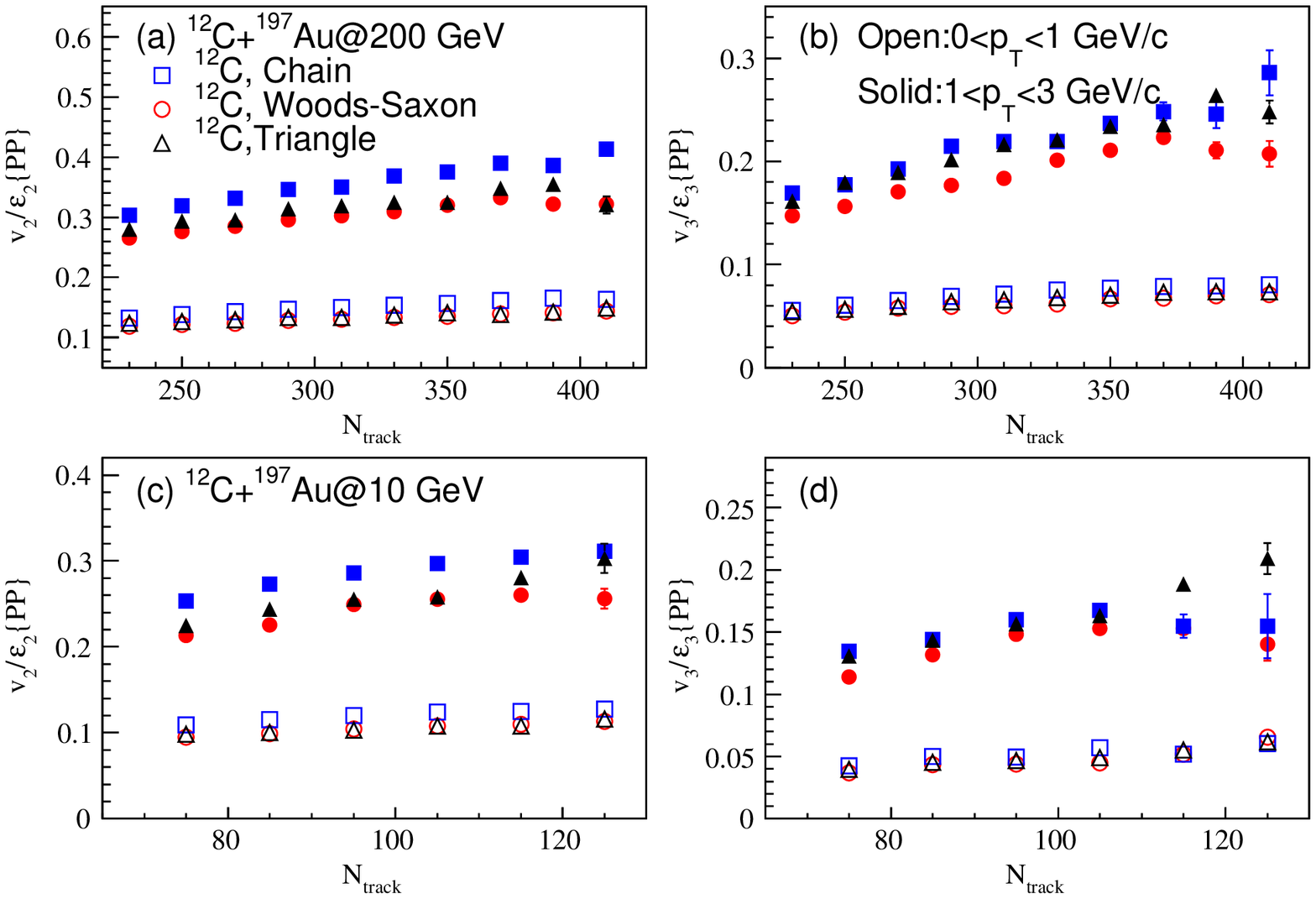}
\caption{
\label{fig:vnOveren}(color online)
In $^{12}\mathrm{C}$+$^{197}\mathrm{Au}$ collisions at $\sqrt{s_{NN}}$ = 200 GeV (upper panels) and 10 GeV (lower panels),  (a) and (c) $v_2/\epsilon_2\{PP\}$,  (b) and (d) $v_3/\epsilon_3\{PP\}$ of final hadrons as function of $\mathrm{N_{track}}$.
 }
\end{figure}

Figure~\ref{fig:v2Evolution} and Figure~\ref{fig:v3Evolution} present elliptic flow ($v_2$) and triangular flow ($v_3$) as  a function of $N_{track}$ in  $^{12}\mathrm{C}$+$^{197}\mathrm{Au}$ collisions at $\sqrt{s_{NN}}$ = 200 GeV and 10 GeV in the AMPT model with different configured pattern of $^{12}\mathrm{C}$. The initial partons inherited from the HIJING~\cite{HIJING-1,HIJING-2} model is considered as the initial stage of the collisions. From Fig.~\ref{fig:v2Evolution} (a), (e) and Fig.~\ref{fig:v3Evolution} (a), (e), there is no elliptic flow $v_2$ and triangular flow $v_3$ in the initial stage of the collision system at both  $\sqrt{s_{NN}}$ = 200 GeV and   10 GeV. The initial partons will interact each other and then reach freeze-out status, which is simulated by the ZPC model~\cite{ZPCModel}. After the parton interaction simulation in the ZPC model, $v_2$ and t$v_3$ are formed as shown in Fig.~\ref{fig:v2Evolution} (b) and (f), and Fig.~\ref{fig:v3Evolution} (b) and (f). It is obvious that the developed elliptic flow is larger for the Chain structure than for other patterns of $^{12}\mathrm{C}$ and the developed triangular flow  is more significant for the Triangle structure than for other patterns of $^{12}\mathrm{C}$ at both  $\sqrt{s_{NN}}$ = 200 GeV and   10 GeV. This is consistent with the view from hydrodynamical theory~\cite{FlowHydro-1,FlowHydro-2,Song} that collective flow results from the initial coordinate asymmetry. The freeze-out partons are converted to hadrons through a naive coalescence model~\cite{AMPT2005} and here these hadrons do not yet participate hadronic rescattering. Figure~\ref{fig:v2Evolution} (c), (g) and Figure~\ref{fig:v3Evolution} (c), (g) show elliptic flow and triangular flow of hadrons without hadronic rescattering. The formed hadrons will experience hadronic interaction simulated by the ART~\cite{ARTModel} model. The elliptic flow and triangular flow of hadrons with hadronic rescattering are shown in panel (d) and (h) in Fig.~\ref{fig:v2Evolution} and Fig.~\ref{fig:v3Evolution}, respectively. It is seen that there is no significant contribution on collective flow from hadronic rescattering by comparing the flows of hadrons with and without hadronic rescattering. The collective flow of hadrons with/without hadronic rescattering is larger than that of freeze-out parton because of the coalescence mechanism, which brings out the number of constituent quark (NCQ) scaling of collective flow~\cite{NCQJTian,AMPTInitFluc-3} and also NCQ-scaling of nuclear modification factor $R_{cp}$ of hadron~\cite{NCQRcp}. Here the calculation are performed in 
a  transverse momentum window of $0<p_T<1.5\text{~GeV/c}$ for partons and $0<p_T<3\text{~GeV/c}$ for hadrons, with a mid-rapidity window $-1<y<1$. Since we mainly discuss relationship between initial geometry distribution and final momentum distribution, eccentricity coefficients are calculated by using initial partons inherited from the HIJING~\cite{HIJING-1,HIJING-2} and collective flow from hadrons with hadronic rescattering.

\begin{figure}[htbp]
\includegraphics[width=8.6cm]{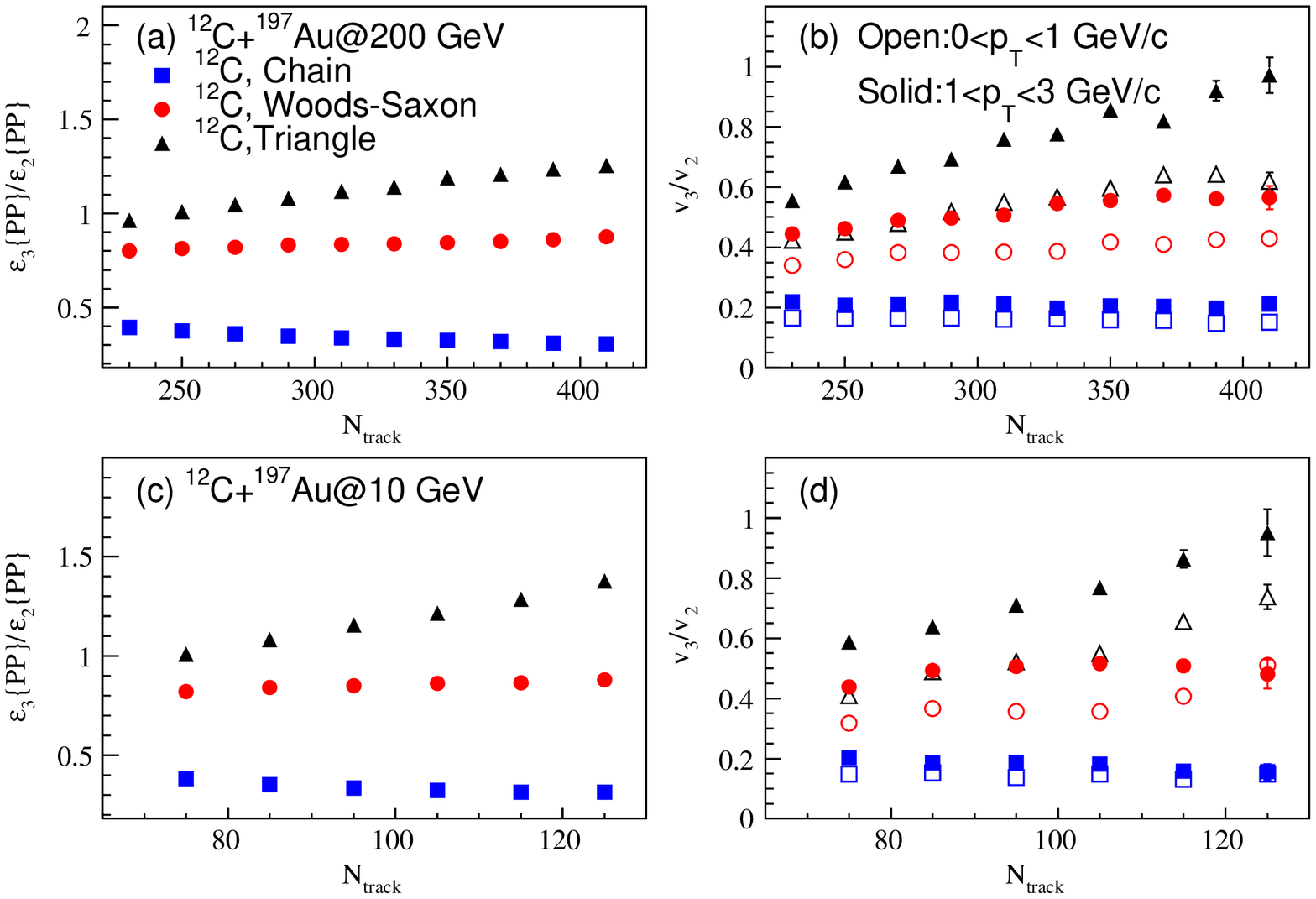}
\caption{
\label{fig:epsilonVn_Ratio}(color online)
 In $^{12}\mathrm{C}$+$^{197}\mathrm{Au}$ collisions at $\sqrt{s_{NN}}$ =  200 GeV (upper panels) and 10 GeV (lower panels), (a) and (c)  $\epsilon_3\{PP\}/\epsilon_2\{PP\}$ , (b) and (d) $v_3/v_2$ of final hadrons as function of $\mathrm{N_{track}}$.
 }
\end{figure}

Now we move to discuss the initial geometry effect on collective flows. From the above discussion, we can only conclude that the Chain structure of $\alpha$-clusters mainly contributes on elliptic flow $v_2$ and while the Triangle structure significantly contributes on triangular flow $v_3$. It will be useful for experimental analysis  if there is a probe to distinguish geometry pattern of $^{12}\mathrm{C}$ through collective flow measurement. In Ref.~\cite{AlphaClusterHIC-1,AlphaClusterHIC-2}, the collective flow ratio between the flow from four-particle cumulant and the flow from two-particle cumulant is proposed as a signature of $\alpha$-clustered $^{12}\mathrm{C}$. In this work another observable will be presented to distinguish the geometry pattern of $^{12}\mathrm{C}$. 

Figure~\ref{fig:epsilonVn_cent} presents the second and third order participant eccentricity coefficients, namely $\epsilon_2\{PP\}$ and $\epsilon_3\{PP\}$, as a function of $N_{track}$ in  $^{12}\mathrm{C}$ + $^{197}\mathrm{Au}$ collisions at $\sqrt{s_{NN}}$ = 200 GeV and 10 GeV in the AMPT model with different configuration pattern of $^{12}\mathrm{C}$.  $\epsilon_2\{PP\}$ slightly increases for the Chain structure pattern of $^{12}\mathrm{C}$ and decreases for the other patterns, with the increasing of $N_{track}$. So does for $\epsilon_3\{PP\}$. The $N_{track}$ dependence of $\epsilon_2\{PP\}$ and $\epsilon_3\{PP\}$ is similar at $\sqrt{s_{NN}}$ = 200 GeV as well as 10 GeV.

Hydrodynamical calculation suggested how the initial geometry distribution (fluctuation) transforms into final flow observables in a heavy-ion collision~\cite{Song,HydroKN-1,HydroKN-2}. And the relationship between initial geometry eccentricity coefficients and flow coefficients are suggested as,
\begin{eqnarray}
v_n = \kappa_n\epsilon_n,  n = 2, 3.
\label{vnknen}
\end{eqnarray}
The response coefficients $\kappa_n$ show the efficiency of the transformation from initial geometry distribution to momentum space in collisions. As discussed in Ref.~\cite{AlphaClusterHIC-1}, $\kappa_n$ depends on the details of collision system and model, the linearity of Eq.~(\ref{vnknen}) allows for model-independent studies for collective flow.

Fig.~\ref{fig:vnOveren} presents $v_2/\epsilon_2$ and $v_3/\epsilon_3$ in panel (a), (b) and (c), (d) at $\sqrt{s_{NN}}$ = 200 and 10 GeV, respectively. The ratio of  $v_n/\epsilon_n$ (n=2,3) increases with  $N_{track}$ in different $p_T$ range and collisions energy. In low $p_T$ range ($0<p_T<1\text{~GeV/c}$), $v_n/\epsilon_n$ (n=2,3) nearly keep  in same line  for the Triangle structure, Chain structure and Woods-Saxon distributions, and show almost independent of $N_{track}$.  In other words, there exists the same efficiency of the transformation from initial geometry distribution to momentum space for the three different  configuration structures of $^{12}\mathrm{C}$ colliding against $^{197}\mathrm{Au}$ in low $p_T$ range. In higher $p_T$ range ($1<p_T< 3\text{~GeV/c}$), however,  there are little difference of  $v_n/\epsilon_n$ (n=2,3) among different cases of configuration structures  of $^{12}\mathrm{C}$, the Chain structure shows a little higher value than other cases. Also, all  $v_n/\epsilon_n$ (n=2,3) display a slight increasing with the $N_{track}$. Therefore, for higher $p_T$ hadrons, 
there are different transformation efficiency from initial phase space to final state for the three different configurations.

From above discussion, we know that the initial geometry asymmetry can be transported to momentum space and remains observed by elliptic and triangular flows. Figure~\ref{fig:epsilonVn_Ratio} (a) and (c) show the ratio of $\epsilon_3\{PP\}/\epsilon_2\{PP\}$ as a function of $N_{track}$ in $^{12}\mathrm{C}$+$^{197}\mathrm{Au}$ collisions at $\sqrt{s_{NN}}$ = 200 and 10 GeV, respectively,  with different configured pattern of $^{12}\mathrm{C}$. The values of $\epsilon_3\{PP\}/\epsilon_2\{PP\}$  show a drop according to the order of the Triangle structure, Woods-Saxon and Chain structure pattern of $^{12}\mathrm{C}$. It is interesting that the ratios of $v_3/v_2$ of final hadrons take the similar order as $\epsilon_3\{PP\}/\epsilon_2\{PP\}$, as shown in Fig.~\ref{fig:epsilonVn_Ratio} (b) and (d). $\epsilon_3\{PP\}/\epsilon_2\{PP\}$  and $v_3/v_2$ all show increasing dependent tend with the increasing of $N_{track}$ in Triangle structure pattern of $^{12}\mathrm{C}$ and approximately keep flat in the other two patterns not only at low $p_T$ but also in higher $p_T$ region.

Therefore the ratio of $v_3/v_2$ is proposed as a probe to distinguish the $\alpha$-clustering structure pattern of $^{12}\mathrm{C}$ in experiment. Especially, we can consider  a nucleus without exotic structure near $^{12}\mathrm{C}$ nucleus  or an isobar nucleus of $^{12}\mathrm{C}$  to collide against heavy-ion as  a reference system. If $v_3/v_2$ in $^{12}\mathrm{C}$+$^{197}\mathrm{Au}$ collisions is obviously larger than that in the reference system, $^{12}\mathrm{C}$ could be constructed by the  three $\alpha$-clusters Triangle structure. If $v_3/v_2$ in $^{12}\mathrm{C}$+$^{197}\mathrm{Au}$ collision is significantly smaller than that in the reference system, $^{12}\mathrm{C}$ can be constructed from the three $\alpha$-clusters Chain structure. If $v_3/v_2$ in $^{12}\mathrm{C}$+$^{197}\mathrm{Au}$ collision is comparable with that in the reference system, $^{12}\mathrm{C}$  should be a non-exotic structure nucleus. And the sharp increasing trend of $v_3/v_2$ as a function of $N_{track}$ can be seen as a probe to distinguish if $^{12}\mathrm{C}$ is formed in triangle shape with three $\alpha$s or not.

In the above results, collective flows and initial geometry coefficients at $\sqrt{s_{NN}}$ = 200 GeV and 10 GeV are presented. However, we did not discuss too much the energy dependence of these results since there is no significant difference in centrality dependence of the flows and initial geometry coefficients between  $\sqrt{s_{NN}}$ =  200 GeV and 10 GeV. We should note that the multiplicity is higher at $\sqrt{s_{NN}}$ = 200 GeV than at $\sqrt{s_{NN}}$ = 10 GeV and from this point the measurements for collective flows will be more feasible at $\sqrt{s_{NN}}$ = 200 GeV from different methods such as reaction plane method  and Q-cummulent method.

\section{Summary}

In summary, the Chain and Triangle $\alpha$-clustered as well as the Woods-Saxon distribution pattern of $^{12}\mathrm{C}$ colliding against $^{197}\mathrm{Au}$ are calculated at $\sqrt{s_{NN}}$ = 200 GeV and 10 GeV in the AMPT model. The effects on collective flow from initial geometry of nucleon distribution are discussed in this work.  The pattern of Chain arranged $\alpha$-cluster significantly contributes on elliptic flow $v_2$ but the Triangle arranged $\alpha$-cluster pattern mainly enlarges triangular flow $v_3$. This is consistent with the viewpoint from hydrodynamical theory on collective flow that the initial geometry asymmetry will be transformed to momentum space at final state. And $v_3/v_2$ is proposed as a probe to distinguish the pattern of $\alpha$-clustered $^{12}\mathrm{C}$  for experimental analysis.

\vspace{0.5cm}

We are grateful for discussion with Dr. Aihong Tang from BNL. This work was supported in part by  the National Natural Science Foundation of China under contract Nos. 11421505, 11220101005,  11105207, 11275250, 11322547 and U1232206, the Major State Basic Research Development Program in China under Contract No. 2014CB845400, and the Key Research Program of Frontier Sciences of the CAS under Grant No. QYZDJ-SSW-SLH002.



\end{document}